# Disrupting Test Development with AI Assistants

Building the Base of the Test Pyramid with Three AI Coding Assistants

**Vijay Joshi,** *Concora Credit Inc., Beaverton, OR, 97006, USA*

**Iver Band,** *Concora Credit Inc., Beaverton, OR, 97006, USA*

***Abstract*— Recent advancements in large language models, including GPT-4 and its variants, and Generative AI-assisted coding tools like GitHub Copilot[1], ChatGPT[2], and Tabnine[3], have significantly transformed software development. This paper analyzes how these innovations impact productivity and software test development metrics. These tools enable developers to generate complete software programs with minimal human intervention before deployment. However, thorough review and testing by developers are still crucial. Utilizing the Test Pyramid concept[4], which categorizes tests into unit, integration, and end-to-end tests, we evaluate three popular AI coding assistants by generating and comparing unit tests for open-source modules. Our findings show that AI-generated tests are of equivalent quality to original tests, highlighting differences in usage and results among the tools. This research enhances the understanding and capabilities of AI-assistant tools in automated testing.**

*Keywords: Unit Testing, AI-Assistant Tools, Generative AI, ChatGPT, Tabnine, LLMs, GitHub Copilot, Testing Automation, Testing Pyramid.*

Testing is a critical component in defining the quality and success of both product and software development life cycle (SDLC). Its primary role is to ensure that the final product meets the desired standards, functions correctly and delivers a satisfactory user experience. Historically, the SDLC has evolved significantly, transitioning from the rigid, sequential processes of the early Waterfall model to the more dynamic and iterative practices of contemporary agile methodologies.

In contrast, modern SDLC practices such as Agile and extreme programming (XP) advocate for a more flexible, collaborative, and iterative approach. These methodologies emphasize testing integration throughout the development process, ensuring quality is maintained from the earliest stages. This shift has been driven by the need for faster product delivery, greater responsiveness to evolving requirements, and overall enhancement of product quality.

For enterprises, robust testing strategies, advanced tools, and clear standards are essential for delivering high-quality, defect-free products. These practices reduce time-to-market and enhance ROI, making product reliability a key market differentiator.

Recent advancements in generative AI, which use machine learning to generate code and automate tasks, are transforming traditional testing methods, including those defined in the testing pyramid from unit to system and acceptance testing.

This paper will explore how generative AI technologies are transforming the testing landscape within the SDLC. We will assess unit test development methods with AI-assisted tools like GitHub Copilot[1], Tabnine[3], and ChatGPT[2], and analyze how these tools are enhancing testing practices. We used specific versions of underlying LLMs offered by these tools and with the caveat that different combinations of LLMs and front ends may yield different results. By evaluating these advancements, we aim to understand how the adoption of AI-assistant tools is shifting the perception and execution of testing, leading to more reliable, efficient, and innovative software development processes.

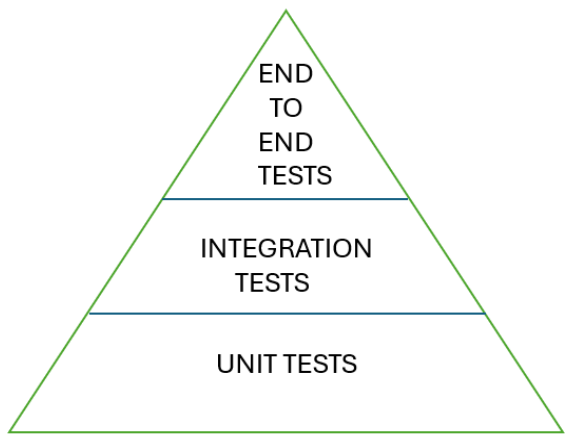

FIGURE 3. The Test Pyramid

# RELATED WORK

In recent years, substantial progress has been made in applying AI to the test automation domain. The following items highlight a selection of related work that captures the intersections between generative AI, AI-assisted tools, and various aspects of testing automation (TA).

In the paper[5], Filippo et al. survey and analyze grey literature to explore the broad adoption of AI in TA, providing insights into AI's current and future roles in the field. However, the paper does not offer a detailed comparative evaluation of popular AI-assisted tools. Instead, it catalogs and identifies 100 AI-driven TA tools. The study underscores that automated test generation and self-healing test scripts are the most common AI solutions.

In the paper[6], Pandey et al examine how GitHub Copilot is transforming software development by evaluating its efficiency and challenges in real-world projects. The study addresses various aspects of software development, including repetitive coding tasks, unit test generation, debugging, and pair programming. Additionally, it assesses GitHub Copilot's benefits and capabilities through its application in real-world projects involving proprietary codebases.

In the paper[7] introduce KAT (Katalon API Testing), a novel approach that utilizes the large language model GPT along with advanced prompting techniques to autonomously generate test cases for validating REST APIs.

The article[8] examine the unit test cases generated by and evaluate the tools Copilot and ChatGPT in a controlled environment. This work is among the closest in the field to analyzing test cases, though it relies on several assumptions and provides a broad analysis of just two tools using different sets of metrics.

The paper[9] compares high-level method generation tasks across four AI-based code assistants: GitHub Copilot, Tabnine, ChatGPT, and Google Bard. The study provides a broad analysis of development methods but does not delve deeply into test development methods and their assessments.

Other work related to testing and utilization of the LLMs technology such as log analysis[10], crash event predictions[11] also has been referred to understand the wide impact of the usage of these AI-Assistant tools.

# CONTEXT

In this paper, our primary focus is to assess the capability of AI-Assisted Engineering tools like ChatGPT, GitHub Copilot and Tabnine on unit test cases generation and established a comparative analysis on original generated test cases against unit test cases generated with the help of these tools. In line with this, we aim to answer several research questions through experimental evaluation.

RQ1: How well these tools are assisting to generate unit test cases?

RQ2: Can these tools provide differentiation between original generated unit test cases and unit test cases generated by these AI-Assistant tools?

RQ3: What are the potential benefits of using these tools in the realm of unit test case generation?

RQ4: Factors impacting the quality of unit test case generation when using these tools.

## METHODOLOGY

- Comparing Unit tests generated with the help of AI-Assisted tools and original generated Unit Test cases using open-source code base.
    - We will generate unit test cases from the open-source repositories using GitHub Copilot, Tabnine, and ChatGPT. We use the following metric for the comparison of unit test cases generated with the help of AI-Assisted tools and original generated unit test cases.
- Coverage of Unit test cases generated as compared to original generated.
    - As compared to original generated unit test cases, observing and analyzing the coverage between two sets of unit test cases.
- Comparing capabilities of AI-Assisted tools in the realm of unit testing.

## SOURCE CODE REPOSITORY

We use the listed open-source code repositories for unit test case generation that help in understanding the detailed nature of our experiments. Then, we perform comparative analysis of the generated unit test cases with the actual ones. Additionally, we use closed source code repositories for the comparative analysis of AI-Assistant tools to perform experiments based on the proposed methodologies.

Junit5 Modular World[12] and Google Java Format[13]

## EXPERIMENTAL SET-UP

For our experiments, we utilize the following AI-assisted tools and configurations:

**ChatGPT 4o mini free version:** Employed both GUI and API access for unit test case generation and prompt engineering for unit test case comparison.

**GitHub Copilot:** Used as an integrated IDE default plugin for unit test case generation and comparison.

**Tabnine Protected and Private Model:** Utilized as an integrated IDE plugin for unit test case generation and comparison.

**Public Open-Source Code Repositories:** Provided access to existing unit test cases and code for generating additional unit tests.

**Closed-Source Code Repositories:** Included for experimentation with proprietary code.

These tools and configurations facilitated our evaluation and comparison of AI-assisted unit test case generation and effectiveness.

## EXPERIMENT DETAILS

**Prompt 1:** Generate unit test case for the "code snippet from the Flavor.java file".
**AI-Assisted Tools:** ChatGPT, GitHub Copilot, and Tabnine.
**Open-source code repository:** Junit5 Modular World[12]

**Prompt 2:** Generate unit test case for the "code snippet from Newlines.java".
**AI-Assisted Tools:** ChatGPT, GitHub Copilot, and Tabnine.
**Open-source code repository:** Google Java Format[13]

**Prompt 3**: Compare unit test cases generated by AI-assisted tools with original generated unit test cases.
<Add original generated test cases from the original test files and newly generated test cases from respective AI-Assisted tools>

**AI-Assisted Tools**: ChatGPT, GitHub Copilot, and Tabnine.

**Open-source code repository:** Junit5 Modular World[12]

**Prompt 4**: Compare unit test cases generated by AI-assisted tools with original generated unit test cases.
<Add original generated test cases from the original test files and newly generated test cases from respective AI-Assisted tools>

**AI-Assisted Tools**: ChatGPT, GitHub Copilot, and Tabnine.

**Open-source code repository:** Google Java Format[13]

NOTE: Same experimentation has been performed upon the closed-source repositories as well. In addition to the above-mentioned prompts. We also used relevant prompts to frame the outcome and response in a way that could be utilized for comparative analysis.

## RESULTS

As we explained in the experimentation details, we categorized the experiments at the following high-level:

Generating unit test cases using the same prompt statement for all three tools.

Comparing the results of unit test cases generated by AI-assisted tools and with the original unit test cases.

The following sections display the results of unit test generation for Flavor. Java and NewLines.java.

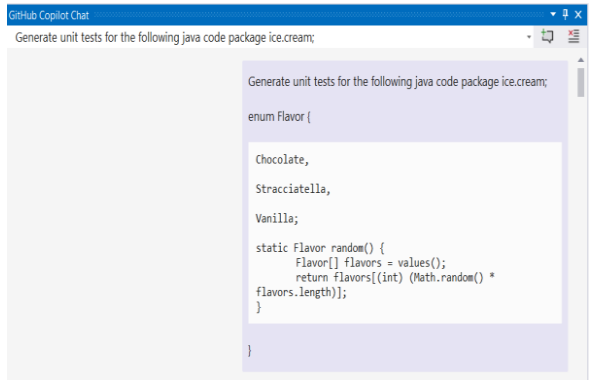

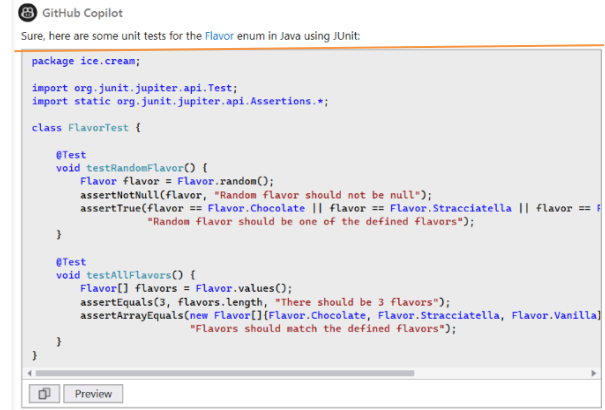

FIGURE 2. GitHub Copilot Chat prompts to generate unit tests for the Flavor.java code.

FIGURE 3. GitHub Copilot Chat generated unit test case for the Flavor.java code.

We used identical prompts with GitHub Copilot, ChatGPT, and Tabnine to generate unit test cases for both open-source and closed-source code. The generated tests showed similar functional behavior and better coverage compared to the originals, though differences in code length and the number of default tests were noted. The AI-generated tests were effective in meeting functional requirements and achieved comparable coverage to the original tests. Simple code required fewer prompts, but complex code needed more detailed prompts for accurate coverage. The tools generated unit test cases with at least 85% coverage for complex scenarios. We compare these results with the original tests in the following sections.

**Comparison: GitHub Copilot versus Original Unit Test Cases**

The original created test file and the one GitHub Copilot generated both aim to test the functionality of the Newlines class, but there are some differences in their structure, style, and coverage. Here are the key differences:

Testing Framework and Assertions:

The original test file uses JUnit4 and Google Truth for assertions, while the generated test file uses JUnit5 and JUnit Jupiter Assertions.

Test Coverage:

The original file tests the 'lineOffsetIterator' and 'lineIterator' methods. The generated file covers a broader range including 'count', 'firstBreak' 'isNewline', 'hasNewlineAt', 'getLineEnding', 'guessLineSeparator', and 'containsBreaks'.

Exception Handling:

The original file uses try-catch and 'fail()' for 'NoSuchElementException'. The generated file uses 'assertThrows'.

Both approaches are valid. The generated test file provides more comprehensive coverage using the modern JUnit5, while the original file, using JUnit4 and Google Truth, focuses on specific iterator methods. This difference reflects the tools available at the time of creation.

**GitHub Copilot - Difference between the coverage of unit test cases**

Original Created Test File:

Focuses on testing the 'lineOffsetIterator' and 'lineIterator' methods.

Provides detailed tests for different newline characters and their positions.

Tests the behavior of iterators when reaching the end of the input.

Generated Test File:

Provides comprehensive coverage of the Newlines class, testing almost all public methods.

Includes tests for counting line breaks, finding the first break, checking for newlines, and more.

Tests the behavior of iterators and other methods with various inputs and edge cases.

The generated test file offers broader coverage of the Newlines class, ensuring that more methods are tested, and more scenarios are considered. The original created test file, while thorough in its specific focus, does not cover as many methods or scenarios.

**Comparison: Tabnine versus Original Generated Unit Test**

The original created test case file you provided is written in Java using JUnit 4 annotations. It tests the 'Newlines' class by checking the offsets and lines of various input strings.

Here's a comparison of the original created test case file with the unit test generated by the code snippet provided earlier:

The original test case file uses the @Test annotation from JUnit 4, 'assertThat' from Google Truth, and 'ImmutableList.copyOf' for iterator conversion. It includes specific methods like 'terminalOffset' and 'terminalLine' to test behaviors with input strings. In contrast, the generated unit test file uses JUnit 5's @Test annotation, 'assertEquals' for assertions, and Guava's 'Iterators.size' and 'Iterators.next' for iterators. It lacks the specific 'terminalOffset' and 'terminalLine' methods found in the original

Overall, the originally created test case file provides a comprehensive set of test cases for the 'Newlines' class, while the generated unit test focuses on specific methods and scenarios.

**Tabnine - Difference between the coverage of unit test cases**

Unit test coverage refers to the extent to which your code is tested by your unit tests. It measures the percentage of your code that is executed during the testing process.

In summary, the original created test case file provides a basic set of test cases for the 'Newlines' class, while the generated unit test provides a more comprehensive and detailed set of test cases. The differences in unit test coverage between the two are primarily due to the number of test methods, assertions, and scenarios covered.

## DISCUSSION

In more complex scenarios, we encountered initial challenges in adopting the right set of prompts. Effective prompt engineering is crucial for generating unit test cases with AI-assisted tools and represents a learning curve for engineers. Incorrectly defining prompts can lead to extended times for generating unit test cases for complex code scenarios.

Conversely, when highlighting specific code sections and asking these tools to generate unit test cases, they often perform better than when extensive prompt engineering is required. In some cases, the tools succeeded in generating accurate unit test cases on the first attempt. For greenfield applications, the tools generally performed well, with many unit test cases providing good coverage and accuracy from the outset.

We relied on the accuracy of the original source code when using the generated unit test cases. We assumed that the original source code was free of bugs with respect to its functional and syntactical behavior. Given the difficulty in verifying the functional accuracy of original source code repositories, we focused on checking its syntactic accuracy and behavior before generating unit test cases. Any flaws in the original source code could potentially lead to inaccuracies in the generated unit test cases.

With a comprehensive set of tests, AI tools offer developers the flexibility to choose which tests to include, modify, or exclude, thereby avoiding unnecessary test bloat. These tools demonstrate their ability to include essential tests by replicating examples that focus on critical aspects. They increase the likelihood of comprehensive code coverage by minimizing the marginal costs of additional tests, though developers must still review and finalize the tests to ensure their relevance and effectiveness with real-world and complex projects.

## CONCLUSION

Several AI-assisted tools are available at the time of writing this paper. We chose to analyze three specific tools based on the following considerations:

**Tabnine** was selected for its strong security, legal, and compliance features, particularly when using proprietary models in a controlled environment. It offers significant compliance credentials like SOC 2 Type 2, GDPR, and ISO 9001, which help safeguard data security and privacy—features not always present in other tools.

**GitHub Copilot** was chosen due to its popularity within the open-source community and its integration with GitHub, the largest open-source code repository. GitHub Copilot benefits from extensive publicly available code used in its training and is a natural choice for organizations already using GitHub as their source code management platform.

**ChatGPT** was included because it is based on the widely recognized GPT models from OpenAI. Its maturity in generating high-quality code and its global popularity make it relevant for comparative analysis.

The analysis of outcomes from these tools revealed interesting insights. Each tool shows the potential to significantly impact test development, though their effectiveness varies across several dimensions. Key factors for further consideration include:

**Functionality and Uses:** Tabnine and GitHub Copilot excel in unit test generation within IDEs, while ChatGPT offers extensive natural language processing capabilities, making it broadly applicable in unit test generation.

**Integration with Enterprise SCM:** GitHub Copilot is particularly beneficial for organizations using GitHub as their SCM (Source Code Management) tool. Tabnine, being IDE-agnostic, integrates well across various environments.

**Security and Privacy:** Tabnine excels in strict security and compliance, addressing concerns like IP protection and data security.

**Cost and Licensing:** Tools vary in pricing and licensing. Some offer free tiers, while others are subscription-based, which organizations should evaluate based on their needs.

**Developer Experience (DX):** The ease of use and fit into development workflows are crucial.

**Support and Documentation:** Quality support and comprehensive documentation are important for effective tool usage.

**Community and Ecosystem:** The size and activity of the user community and integration with other platforms are relevant considerations.

**Customization:** The ability to customize tools to meet specific needs or coding standards.

**Market Stability:** Research the tool's reliability and track record for future updates.

Future research will expand the analysis of the test pyramid, focusing on Integration and End-to-End Testing. We also plan to evaluate AI-generated unit test cases with different LLMs, such as Google Bard, to compare outcomes across models.

Our study provides a foundational understanding of these tools' impact on test development. We aim to guide further research on how these tools can influence broader test development methods within the software development lifecycle.

**Vijay Joshi** is a Senior Enterprise Architect at Concora Credit Inc., Beaverton, OR, USA, a provider for credit programs. Prior to that, he served as an Enterprise Architect, Solutions Architect, IT Engineering Leader, and Project lead for large financial, product engineering, and automotive institutions. He delivered Architecture, and managed product development ranging from small to large scale in terms of market impact. He has an extensive experience in the Portfolio and Program Management, Enterprise and Solutions Architecture, Technology Management and Strategy, Cybersecurity, GRC, Data Architecture, Business Systems design and development of large-scale distributed systems. He received MS degree in the Computer Science and specialization in Artificial Intelligence and Machine Learning from Portland State University.

**Iver Band** is VP, Architecture at Concora Credit, a provider of credit programs for non-prime consumers, where he leads a practice that guides complex technology initiatives. Previously, he was Chief Architect at DocVocate, where he enhanced, scaled, and secured a cloud-native healthcare revenue cycle management solution. Prior to that, he served as an enterprise architect, part-time data scientist, IT director, engineering manager, software architect, and infrastructure engineer for Cambia Health Solutions, Standard Insurance, and Hewlett-Packard, where he led development of a patented method of network security management as the second Visiting Technologist at HP Labs. In 2017, he chaired the Open Group ArchiMate Forum, where he advanced an international standard for enterprise architecture modeling. Iver has presented at many technology conferences, and, through his affiliation with EA Principals, he publishes, trains, and advises on enterprise and solutions architecture for major corporations and government agencies.